\documentclass[fleqn,10pt]{wlscirep}
\usepackage[utf8]{inputenc}
\usepackage[T1]{fontenc}

\title{The Economic Complexity of the Roman Empire}

\author[1,+]{Matteo Mazzamurro}
\author[1,+]{Petra He\v{r}m\'{a}nkov\'{a}}
\author[2,*]{Michele Coscia}
\author[2]{Tom Brughmans}
\affil[1]{Social Resilience Lab, Aarhus University, Jen Chr. Skous Vej 4, Aarhus, DK}
\affil[2]{IT University of Copenhagen, Rued Langgaards Vej 7, Copenhagen, DK}

\affil[*]{To whom correspondence should be addressed: mcos@itu.dk}
\affil[+]{These authors contributed equally to this work.}


\begin{abstract}
Economic complexity is a powerful tool to estimate the productive capabilities and future growth of modern economies. Little is known of how economic complexity evolves over long periods in history. In this paper, we use archaeological evidence from the Roman Empire in the form of short texts preserved on a durable material (i.e. inscriptions) to estimate the economic complexity of the various provinces of the empire. By connecting the occupations listed in the text of inscriptions with the location in which the inscribed objects were found we can estimate that the most complex areas during the first four centuries of the Roman Empire have a remarkable and statistically significant overlap with the most complex countries today. While we lack an explanation for the reason of the preservation of economic complexity through the ages, this evidence provides a suggestion about how difficult the development of economic capabilities might be.
\end{abstract}

\begin{document}

\flushbottom
\maketitle
\thispagestyle{empty}

\section*{Introduction}
Economic complexity has emerged in the literature as a way to estimate a country's productive knowledge \cite{hidalgo2021economic, balland2022new}. Normally estimated as the Economic Complexity Index (ECI) -- of which there are multiple alternative definitions \cite{hidalgo2009building, tacchella2012new, cristelli2013measuring} -- it has gained relevance because it is able to predict not only economic growth \cite{hausmann2014atlas}, but also development trajectories -- which economic activities will appear where next \cite{hidalgo2018principle, coscia2020knowledge, pinheiro2022time}. Other applications involve discussing income inequality \cite{hartmann2017linking}, the green transition \cite{mealy2022economic}, and more \cite{hidalgo2023policy}.

Given how much of our society's current and future challenges can be informed by economic complexity, we should devote some attention to understanding how we can affect it -- specifically how it evolves. While most of the cited works -- and more \cite{zhu2017economic} -- study ECI changes in the short and medium term, no one has had the opportunity to do so on long historical time scales.

In this paper, we propose to take a long term -- millennia-wide -- look at economic complexity. By exploiting archaeological data on the Roman Empire, we can create the most accurate picture to date about its occupational and economic structure. In turn, this picture is instrumental in allowing us to estimate the economic complexity of the empire's provinces, identifying which areas of the empire contributed most to its complexity. Finally, we can connect the ancient economic complexity to the modern one. The Roman Empire was a continental polity, which spanned $36$ modern countries. We have an ECI estimation for each of these modern countries based on contemporary trade data and we can correlate it with the ECI based on archaeological data.

This work is an interdisciplinary effort, putting together network science and archaeology -- a novel approach that has been gaining increasing attention in the past years \cite{brughmans2013thinking, knappett2013network, collar2015networks, peeples2019finding}. 

The performance of the economy of the Mediterranean world under Roman imperial rule has been studied through the quantitative analysis integrating archaeological evidence, such as pottery, coins, and most recently, inscriptions, with economic analysis \cite{bowman_quantifying_2009, brughmans_simulating_2022, ortman_identification_2024, fluckiger_roman_2022}. We build on previous work that reconstructed the occupational profile of the Roman Empire from inscriptions \cite{kavse2022division}. We use the largest digital dataset of Latin inscriptions with over 500,000 records as our main source of evidence \cite{kase_2023_list_1_0,hermankova_inscriptions_2021}. These inscriptions contain mentions of ancient occupations, as well as spatiotemporal data to build the occupational profile of the Roman Empire.

After producing the most accurate to date picture of the occupations-based economic structure of the Roman Empire and estimating the ECI of each province, we find that there is a weak but marginally significant ECI correlation for modern countries, between the index as calculated with archaeological data in Roman imperial times and its version with modern trade data -- data recorded around two millennia later. While we are unable to discern whether this correlation is due to long-term effects of economic complexity or to exogenous factors such as geographical positioning or natural resources \cite{tabash2022dynamic}, this result hints at potential implications for development policies.

Regardless of the causes, relative economic complexity rankings appear to be enduring, meaning that -- without a conscious and possibly global effort -- the income disparities between developed and developing countries might linger indefinitely, as these might be fostered by enduring differences in economic complexity \cite{pinheiro2025dark}.

\section*{Results}

\subsection*{Occupational Structure}
We analyze the digitized textual content of more than $525,870$ Latin inscriptions created during the Roman Empire. Most inscriptions date to the first four centuries CE \cite{kase_2023_list_1_0}, a period that this study focuses on as it covers most of the duration of the Latin-speaking western Roman Empire. Inscriptions are short texts surviving on an object made from a durable material and were created for a variety of purposes: to mark someone's grave and provide details of their life, to celebrate their life achievements and bestow honours, to regulate, to inform, but also to materialize an insult, or joke, or to invoke magical powers. Upon computer-assisted reading of these texts, we discovered $8,475$ inscriptions that mention at least one occupation (from one to $30$ mentions of occupations in a single inscription). Of these, 37\% ($3,116$) are funerary inscriptions, reporting on the deceased person's profession \cite{saller_household_2007}, and 11,4\%  are honorific texts ($952$) typically listing the life achievements of the honourand, including their occupation, known as the \textit{cursus honorum} \cite{Salomies2001}. 

Most inscriptions have a known findspot identified with latitude and longitude coordinates, which allows us to precisely geolocate $10,353$ occurrences of $512$ unique occupations throughout the Empire -- under the assumption that mobility during the Roman Empire was less frequent than nowadays \cite{scheidel2007demography}. Detailed evaluation of ancient mobility using epigraphic sources has shown that most people migrated locally, with only a small group of highly mobile individuals \cite{sobotkova_soldier_nodate}. We provide more information about the raw data in the Materials \& Methods section.

\begin{figure}[t]
\centering
\includegraphics[width=\linewidth]{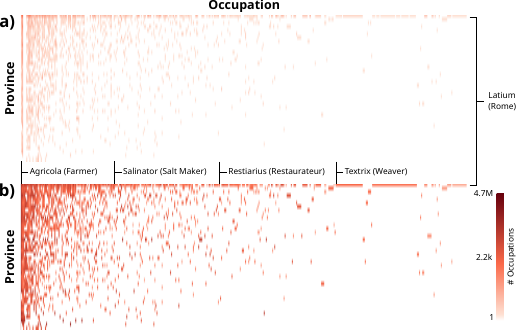}
\caption{(Log) Number of occupations (x axis) in a given province (y axis). (a) Raw counts from inscriptions; (b) bias-corrected counts. The darker the red hue the more occupations were in a given province -- color map consistent across the two figures. The rows and the columns are sorted by their respective sums.}
\label{fig:province-occupation-count}
\end{figure}

We count the number of occurrences of a given occupation per Roman province. We use the Roman province borders from ca. 200 CE digitised by The Ancient World Mapping Centre \cite{ancient_world_mapping_centre_ancient_nodate} and edited by Adam Pažout in 2023). We exclude provinces east of the Euphrates which were only occupied for a very short period, and therefore, we assume, they did not contribute to the overall economic complexity of the Roman Empire. All boundaries were clipped to modern coastlines. Inclusion of Italian \textit{regiones} enhances granularity of the results in the Italian peninsula. Unfortunately, a similar finer division is not entirely possible for other provinces. We only include provinces with at least $1,000$ inscriptions in Latin to ensure the data is representative of the past provincial situation. Figure \ref{fig:province-occupation-count}(a) shows the result.

For our main results, we only use Latin inscriptions, as Latin was the epigraphic language of the Roman Empire, with most surviving datapoints. We also have access to Greek inscriptions \cite{kase_2023_gist_1_1}, which come predominantly from the Eastern part of the Roman Empire, see Materials \& Methods section. Due to the current format of the digital dataset, the structurally different character of the Greek language and the non-standard epigraphic markup do not allow for the application of the same methods as in Latin inscriptions, and therefore, it cannot be used to inform our main results. However, we show in Supplementary Materials Section 2 how including the Greek inscriptions with some assumptions about their content might lead to similar results as the ones described here.

Given that we work with archaeological data, there are a number of biases that would lead these counts not to be accurate. We identify several sources of biases -- language of publication, gender, status, and amount of research dedicated to different areas. We detail how we dealt with these sources of bias in the Supplementary Materials Section 1. 

Our bias correction results in a probabilistic count of the number of occupations in a given Roman province. Figure \ref{fig:province-occupation-count}(b) shows the resulting matrix, aggregated by taking the average of all counts. It represents the most accurate to-date representation of the occupation-based structure of the Roman Empire, counting how many people were involved in a given occupation in a given province. It takes the form of a $M_{po}$ matrix, with $p$ being a province and $o$ being an occupation. Given the high skewedness in the occupation counts, we always operate on the logarithm of the values in $M_{po}$ -- for both visualization and analysis, as is standard practice in economic complexity literature \cite{hidalgo2021economic}.

$M_{po}$ has a clear nested structure. Nestedness is a phenomenon originally observed in ecosystems \cite{patterson1986nested, atmar1993measure}, which has been widely observed in virtually any economic system \cite{hausmann2011network, bustos2012dynamics, neffke2013skill, hidalgo2021economic, balland2022new} -- when analyzed using the same approach we employ here (counting economic activities in geographical places).

Nestedness can be estimated by comparing the $M_{po}$ density on the left and on the right from its isocline -- the line dividing the matrix into its dense upper-triangular part and the rest of the matrix. The bias-corrected $M_{po}$ is more than five times denser on the left of the isocline than on the right. More precisely, it is $5.7$ times denser, a value that is statistically significant ($p < 0.01$) when compared to $1,000$ randomizations of $M_{po}$ where we generate digital twins of each Roman province (preserving its number of occupations and occupation diversity). Supplementary Material Section 3 contains details on the calculation of the isocline and the comparison with a null model.

\begin{figure}[t]
\centering
\includegraphics[width=0.9\linewidth]{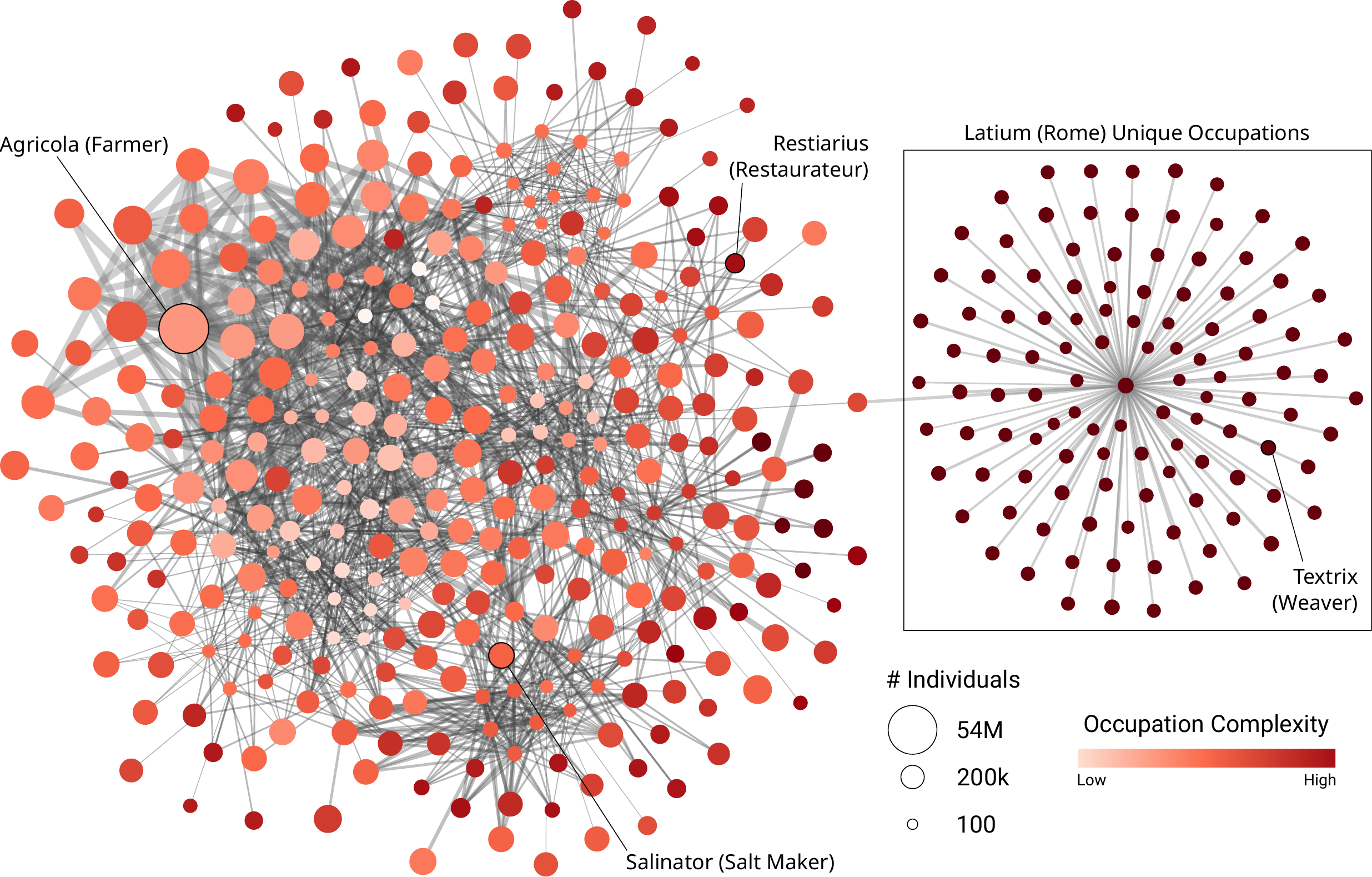}
\caption{The Occupation Space of the Roman Empire. Each node is an occupation. Two occupations are connected if they co-appear in significant numbers in the same provinces. The node size is proportional to the number of people in that occupation. The node color is proportional to the complexity of the occupation from dark red (complex occupation) to white (simple occupation). The edge thickness is proportional to the number of co-appearances, while color and transparency are proportional to the statistical significance.}
\label{fig:occupation-space}
\end{figure}

Following the literature on economic complexity \cite{hidalgo2007product, hausmann2014atlas}, in Figure \ref{fig:occupation-space} we build the Occupation Space of the Roman Empire, connecting occupations if they co-appear in the same provinces a significant amount of times -- more details in the Materials \& Methods section. Such co-appearance is a strong indication of both horizontal and vertical economic integration between the occupations -- they either require similar skills or are part of interconnected supply chains.

The Occupation Space has some evident features. The large star on the right includes occupations only found in Latium (an area around the capital Rome). The denser main part of the Occupation Space has a noticeable left-to-right color gradient, moving from least to most complex occupation, just like the modern Product Space \cite{hausmann2014atlas}. The internal structure, just like the Product Space, shows some clusters, with a main core on the left including common occupations such as \textit{agricola} (farmer), \textit{curator} (manager), and \textit{faber tignarius} (carpenter); and smaller communities on the top and bottom right.

We can exploit the structure of the Occupation Space to estimate the similarity between the Roman provinces. Rather than simply counting the number of occupations they have in common, we can use network distances to weight common occupations according to how close they are to each other in the Occupation Space \cite{coscia2020generalized} -- more details in Supplementary Materials Section 4.

\begin{figure}[t]
\centering
\includegraphics[width=0.9\linewidth]{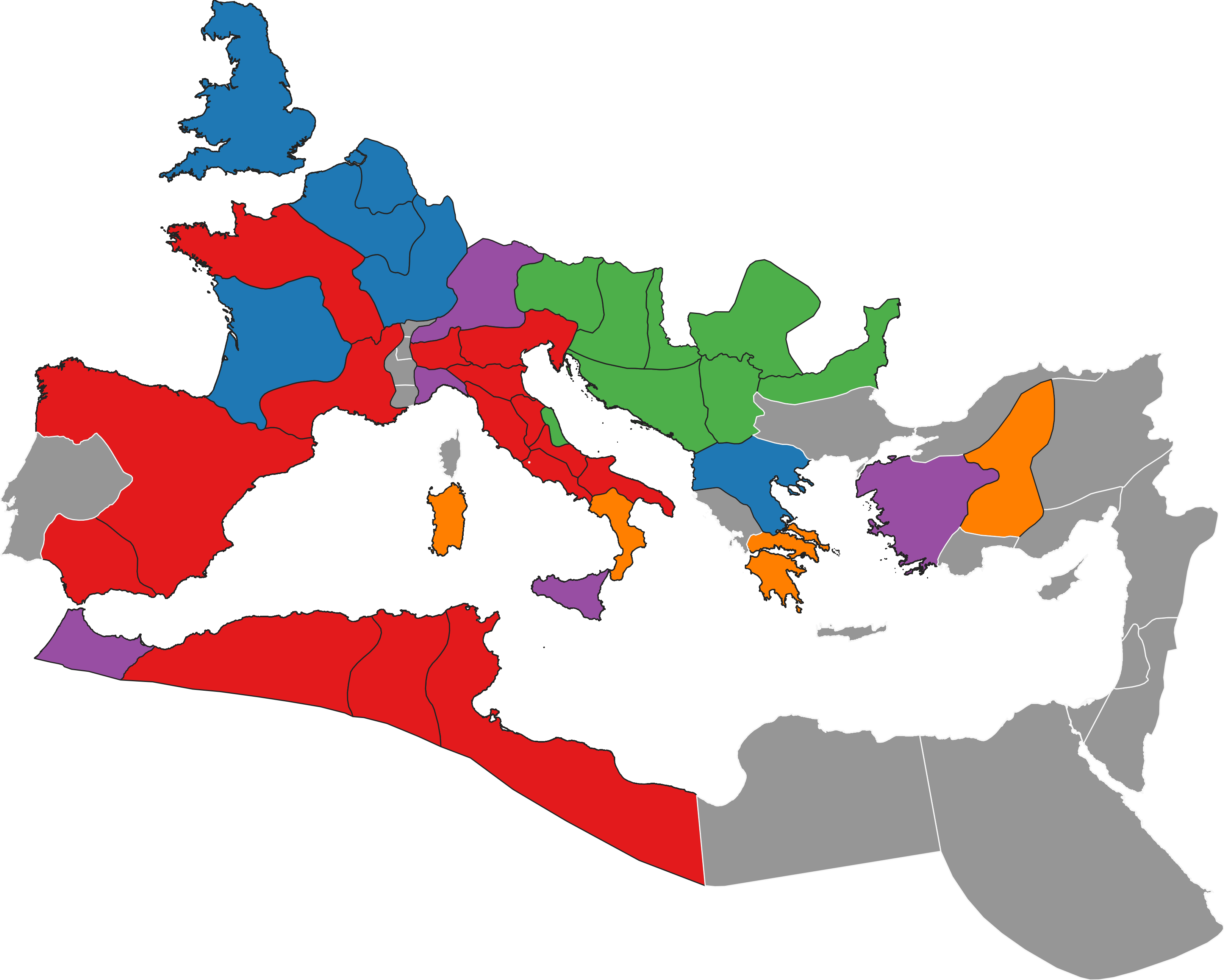}
\caption{The province clusters on the Occupation Space. Provinces in the same color belong to the same cluster. In gray we have provinces with either no occupation data or outliers that cannot be clustered in clusters larger than three elements.}
\label{fig:province-clusters}
\end{figure}

Once we have a province-province distance measure on the Occupation Space, we can analyze all pairwise distances to find clusters of provinces \cite{damstrup2023unsupervised} -- which we show in Figure \ref{fig:province-clusters}. Provinces are grouped together if they have a similar prevalence of occupations that are nearby in the Occupation Space. The purple, orange and blue clusters group together diverse provinces throughout the empire, whereas the green cluster demonstrates that the provinces along the Danube had more similarity in the documented occupations due to the enhanced military presence in the region. The red cluster reveals similarity in occupations between Italian \textit{regiones} and major agricultural-surplus-producing regions active in long-distance trade with a high ECI score (see next section). We describe the clustering procedure in more detail in the Supplementary Materials Section 4.

\subsection*{Economic Complexity}
We can exploit the nestedness of $M_{po}$ to calculate the Economic Complexity Index (ECI) of the provinces of the Roman Empire. We use the classical definition of ECI, where a province is ranked as highly complex if it possesses occupations that are only rarely present in the empire. If a province only hosts occupations that are present in many other provinces, it instead has low complexity. We detail the computation of ECI in the Materials \& Methods section, as well as an argument as to why we do not consider alternative definitions of ECI \cite{tacchella2012new}.

\begin{figure}[t]
\centering
\includegraphics[width=0.9\linewidth]{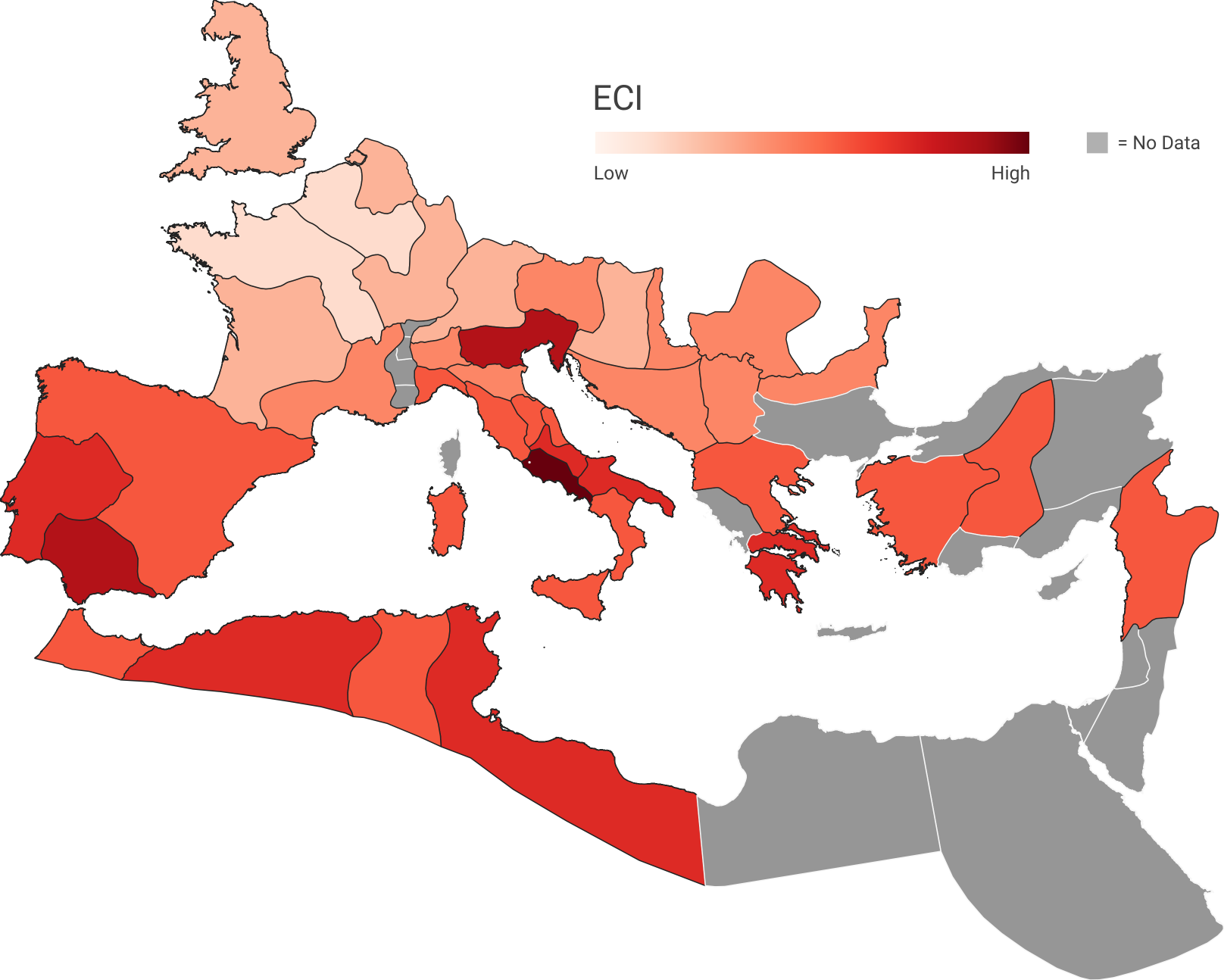}
\caption{The ECI values (color) per province. Darker provinces have higher ECIs. Gray for provinces without an ECI estimate.}
\label{fig:eci-avg-by-province}
\end{figure}

Figure \ref{fig:eci-avg-by-province} shows the result. As expected, Latium and Rome have by far the highest economic complexity. In general, economic complexity is higher in provinces containing key long-distance trading nodes such as Venetia, Carthage, and Baetica. This suggests that trading and exchanges were important to create a strong ecosystem of occupations.

\subsection*{ECI Temporal Stability}
We now turn to study the stability of the Economic Complexity rankings over the course of Western history. Note that we take a correlational approach: we do not investigate what causes an ECI ranking, only that it is somewhat stable over the course of a millennium and a half. Establishing whether this stability is endogenous -- complexity preserves itself -- or exogenous -- complexity comes from external factors such as the geographical position of a trade node -- is left as a future work.

We calculate the Spearman rank correlation coefficient between the ECI rankings of modern countries both using our ancient archaeological data and data coming from UN Comtrade for the period 1962-2022 \cite{growth2019international}. We use a rank correlation as the specific ECI values are unimportant and the distribution of the values from the Latin inscriptions is skewed, with Italy being a far outlier that could drive the correlation.

We obtain a correlation of $0.439$ with $p < 0.05$, indicating that there is a significant positive correlation between the complexities of a country when calculated with data from the Roman Empire and data from modern international trade. The fact that we are able to find a significant correlation with only $23$ observations is remarkable, considering that the two datasets are 1,500 years apart. This result is robust to the introduction of the Greek inscriptions, as we show in Supplementary Material Section 2. 

\begin{table}[]
\centering
\begin{tabular}{r|ll}
ECI Rank & Ancient & Modern \\
\hline
1 &  Italy            & Germany\\
2 &  France           & Switzerland\\
3 &  Germany          & \textbf{Austria}\\
4 &  Spain            & \textbf{United Kingdom}\\
5 &  Belgium          & Slovenia\\
6 &  \textbf{Algeria} & Hungary\\
7 &  Slovenia         & Italy\\
8 &  Hungary          & France\\
9 &  \textbf{Serbia}  & Belgium\\
10 & Switzerland      & Spain\\
\end{tabular}
\caption{The top ten most complex countries out of the 23 countries covered by the inscriptions, according to the ECI calculation based on archaeological data (left) and on modern trade (right). In bold we align the disagreements in the top ten.}
\label{tab:eci-ranks}
\end{table}

Table \ref{tab:eci-ranks} shows the top ten countries with the highest ECI values in both the ancient and the modern data. As we can see, there is substantial agreement: $8$ out of $10$ countries are present in both top tens. Algeria and Serbia were more prominent in ancient times, while data from the Roman Empire failed to capture the rise of the United Kingdom and Austria over the centuries.

\section*{Discussion}
In this paper, we use digitized geolocated inscription data to create the most accurate up-to-date picture of the occupational structure of the Roman Empire. By recording how many inscriptions report the profession of an individual in a given Roman province, we are able to estimate the economic complexity of different areas of the empire. After correcting for a number of biases, we are able to correlate the ancient economic complexity with the one estimated in modern times by using international trade data.

Our results suggest that economic complexity preserves over long historic time spans: the complexity as inferred from inscriptions dating mostly to the 1st-4th centuries CE significantly correlates with the one estimated with 1962-2022 trade data. Our study is correlational in nature, therefore we cannot explain the reasons for this historical resilience in economic complexity. However, by finding evidence for it we have now uncovered an interesting puzzle. Explaining this unexpected resilience can open new research pathways, which can be useful to understand how we increase economic complexity today -- potentially narrowing down the knowledge and economic gaps, and support developing countries.

While interesting, there are a number of factors that need to be taken into account to contextualize our results. First and foremost, when dealing with archaeological data, we need to be aware we are looking at an extremely biased sample. We can only analyze the inscriptions that survived up until now, which is the definition of survivorship bias. Moreover, many inscriptions that could have been produced were not, because the authors themselves were biased -- against the epigraphic representation of women, lower status individuals, and so on.

Crucially, even if the inscriptions were produced and survived, still more biases creep in. In order to be included in our dataset, some researchers must have worked on them, and researchers have biases as well -- focusing more or less on different areas, cultures, time periods, or simply lacking the financial resources to process all the information. We can see this issue directly in our work: we have the coded occupations for Latin inscriptions, but not for the Greek ones.

We have detailed in the paper our strategies to deal with these biases -- showing how ECI rankings are robust to our best guesses on the upper and lower bounds of the occupation counts for provinces. However, our results will for sure become more accurate and robust with additional research. The most natural first step would be to properly codify the occupations from the Greek inscriptions so that we do not have to guess their content by looking at the Latin ones.

There are a number of future research paths opened by this paper. The most important of them is the attempt to explain the reasons for the preservation of economic complexity over the ages. Is this because of exogenous factors -- e.g. the location of a country determines its economic structure, as Italy will always be a crucial trade node in the Mediterranean --, or endogenous -- i.e. being already complex gives a strong head-start to develop even more complexity in the future? Other interdisciplinary future work will involve economists: can we use what we learned about the difficulties in changing the economic landscape to foster economic growth and narrow down income inequalities across countries?

\section*{Materials and Methods}

\subsection*{Data}

\subsubsection*{Inscriptions}
For the main results of the paper, we rely on the \textit{Latin Inscriptions in Space and Time} ($L_{ALL}$) dataset \cite{kase_2023_list_1_0, kase_2024_list_1_2}. This dataset includes $525,870$ Roman inscriptions in Latin, with their geographic coordinates, dates and type -- e.g. epitaph, votive inscription, honorific inscription, etc., along with other attributes. The process leading to the creation of the dataset and the challenges of using the epigraphic data are discussed elsewhere \cite{hermankova_inscriptions_2021}.

In the Supplementary Materials, we test the robustness of our result by also leveraging the information from another dataset: \textit{Greek Inscriptions in Space and Time} ($G_{ALL}$) \cite{kase_2022_gist_0_1, kase_2023_gist_1_1}. This is a smaller dataset, including $217,863$ inscriptions in Greek, with their geographic coordinates and dates. The dataset is not as well-curated as the Latin one, and the distant reading methods allowing for quantified analysis of Greek inscriptions are only being developed.  

\begin{figure}[t]
\centering
\includegraphics[width=.9\linewidth]{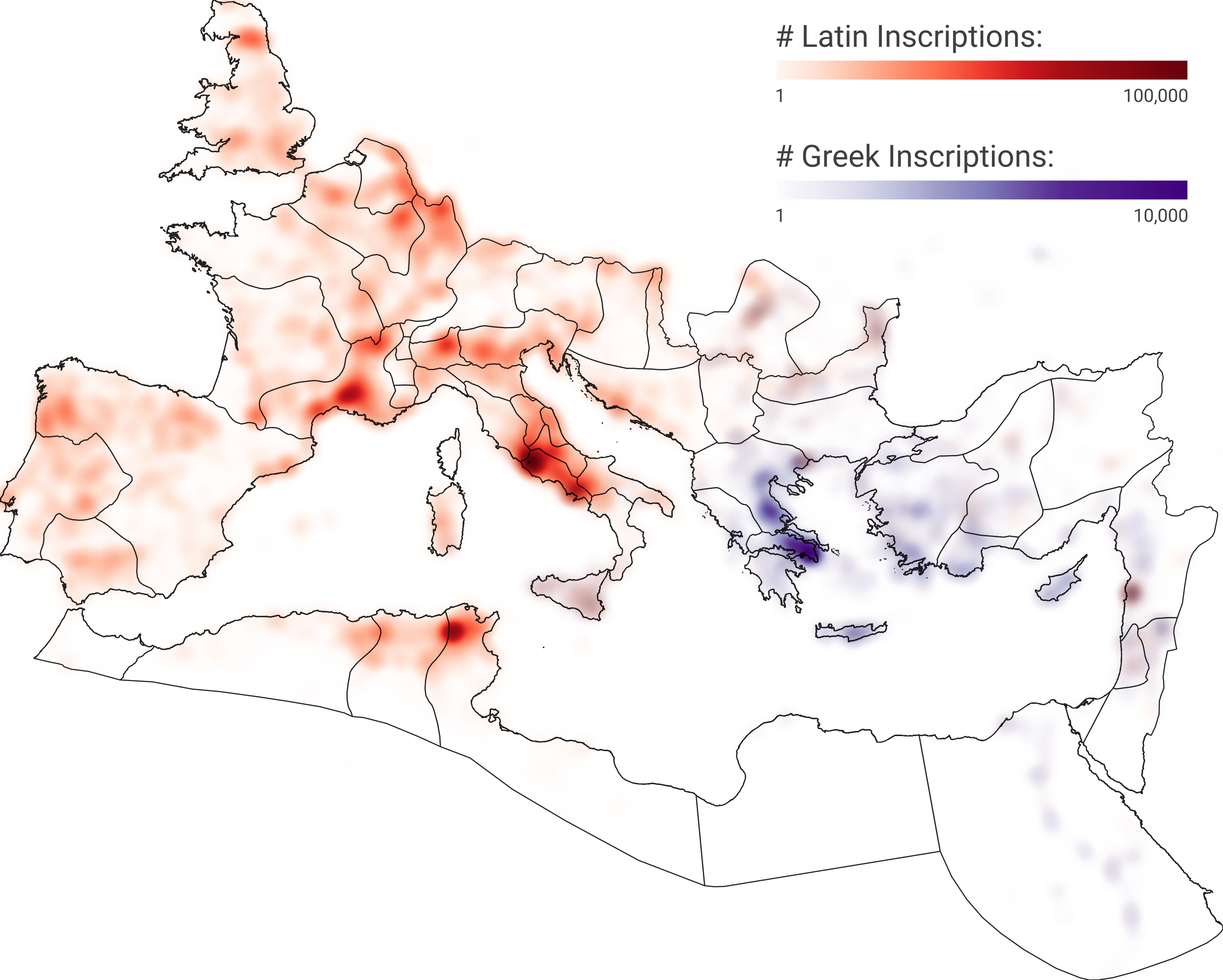}
\caption{The total number of inscriptions in $L_{ALL}$ (red) and $G_{ALL}$ (purple).}
\label{fig:inscription-totals}
\end{figure}

Figure \ref{fig:inscription-totals} shows the geographical coverage of these two datasets. As expected, the Latin inscriptions dominate in the Western part of the Roman Empire, and the Greek inscriptions dominate in the Eastern part.

\begin{figure}[t]
\centering
\includegraphics[width=.5\linewidth]{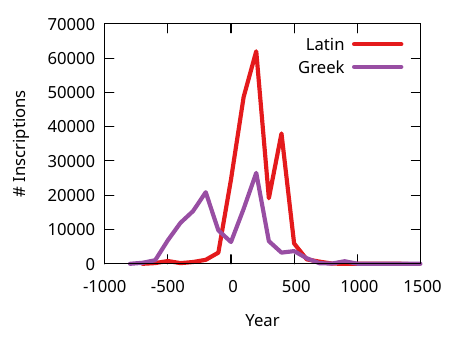}
\caption{The number of inscriptions (y axis) found in a given century (x axis), in Latin (red)  and Greek (purple).}
\label{fig:inscriptions-temporal-spread}
\end{figure}

There are two key differences between $L_{ALL}$ and $G_{ALL}$ that lead us to use only $L_{ALL}$ for the main results and $G_{ALL}$ only for robustness. First, the two datasets span different time periods and with a different temporal spread -- as Figure \ref{fig:inscriptions-temporal-spread} shows. Only about half of the Greek dataset dates to the Roman Empire, with the other half coming from the 6th to 1st centuries BCE. Merging the two datasets therefore is not straightforward and needs caution.

Moreover, the $L_{ALL}$ is structurally richer, as it has received more research attention and curation. Specifically, the $L_{ALL}$ data set has been pre-processed by 
\cite{kavse2022division, hermankova_inscriptions_2021, kase_classifying_2021} to obtain the occupations ($L_{OCC}$), epitaphs ($L_{EPI}$), and people sub-datasets ($L_{PEO}$) -- Supplementary Material Section 1 provides more details about these subsets.

This is not true for the $G_{ALL}$ dataset. Crucially for our work, we cannot reliably extract a $G_{OCC}$ dataset comparable to $L_{OCC}$. The application of current NLP methods and their verification by domain-specialist showed a high level of false positivity in the detection method. Specifically, the implementation of the latest AgiLe Lemmatizer designed for Greek epigraphic data \cite{de_graaf_agile_2022} showed that more than 50\% of occupations were detected in inscriptions where no occupation was mentioned. The results were verified by close reading of a sample of inscriptions by a human domain specialist in epigraphy (=He\v{r}m\'{a}nkov\'{a}). Until the computer-assisted detection and processing methods for Greek are improved, we, therefore, focus on Latin as the main dataset and use Greek to inform the analysis. We detail in Supplementary Material Section 1 how we address this issue in our robustness analysis.

Other datasets used to mitigate the bias in inscription-based data include the Roman cities locations and populations, assembled by Hanson \cite{hanson2019poparticle, hanson2019popdata}, and the population in 10 macroregions of the Empire (Table 3.1 in \cite{scheidel2007demography}), manually associated with the provinces that constitute them.

\subsubsection*{Modern Data}
To calculate the modern ECI values, we rely on the UN Comtrade dataset as made available in the Harvard Dataverse \cite{growth2019international}. We use the SITC Rev 2 dataset because we want to have as wide a time interval as possible. Shorter intervals might be dominated by short-term fluctuations in global trade, which would fail to be representative of the long-term economic structures we can capture over the course of history. By calculating the ECI with data spanning decades we are more likely to have a representative estimate.

The SITC Rev 2 data spans from 1962 to 2022, while the HS data only starts in 1995 and is therefore discarded. We calculate the ECI for this data with the exact same methodology as the one used for the ancient ECI, which is detailed later on in this section.

To calculate the modern ECI, we select a subset of countries and products. We remove all country codes not referring to actual countries, countries with less than 300k inhabitants, lower than 300 million USD GDP, or that represent less than $0.05\%$ of the trade volume (combined). At the same time, we drop all products which taken together represent less than $2.5\%$ of the world's trade -- in practice, we preserve $97.5\%$ of the total trade volume. This is done because the ECI calculation can become sensitive to extremely small countries and products.

\subsection*{Bias Mitigation}
Working with archaeological data always implies dealing with a number of biases in the data. For this work, we identified five main sources of bias. We briefly outline our strategies to deal with them here and provide further details in the Supplementary Materials Section 1.

In general, we identify reasonable bounds for our corrections, and then we draw random samples from these distributions, leading to many different plausible $M_{po}$ matrices.

\textbf{Status}. Prestigious professions and occupations elevating social status tend to be over-represented in inscriptions. At the same time, the low-prestige jobs performed mostly by slaves and household servants remain unrepresented in epigraphic data proportional to the expected demographic composition of the society \cite{joshel_work_1993, treggiari_jobs_1975}. To counteract status bias, we inflate the counts of non-elite occupations. For the inflation factor, we draw from different distributions assuming an elite prevalence between 1\% and 5\% of the population, as suggested in~\cite{scheidel2009size}.

\textbf{Gender}. Men are disproportionately mentioned in inscriptions compared to women, regardless of their status, with roughly two men mentioned per woman. This gender unbalance is well documented in research \cite{saller_household_2007, joshel_work_1993, treggiari_jobs_1975, scheidel_most_1995, scheidel_most_1996, boserup_womans_1970, becker_roman_2016, treggiari_lower_1979}. For this reason, we assume we need to double the number of women in our data, drawing from a distribution bounded by our best estimate about what the minimum and maximum gender disparity should be.

\textbf{Primary Sector}. Agricultural (and other primary sector) work is absent in the inscriptions, despite it occupying roughly 85\% of the population \cite{scheidel_most_1995, bowman_roman_2013}. We then inflate the count of the generic profession of farmer, with a proportion roughly similar to that of the rural population.

\textbf{Research}. The number of inscriptions per region is not proportionate to their historical population, reflecting not only cultural differences but also the state of investigation within these regions \cite{bodel_epigraphic_2024, mullen_latinization_2024, woolf_monumental_1996, meyer_epigraphy_2011}. We attempt to mitigate this issue by normalizing the occupation counts to reflect the population size of each province.

The final source of bias only applies to the robustness tests in the Supplementary Material:

\textbf{Language bias}. The data set only contains inscriptions in Latin, but Greek was an important language of epigraphic publication, especially in the Eastern Roman Empire and some parts of the Western Roman Empire \cite{prag_epigraphy_2013}. Extracting occupation-related data from Greek inscriptions has not yet been implemented for technical reasons. Thus, we need to estimate the occupational data contained in them by inflating the occupation counts from the Latin inscriptions with the number of found Greek inscriptions.

\begin{figure}[t]
\centering
\includegraphics[width=0.45\linewidth]{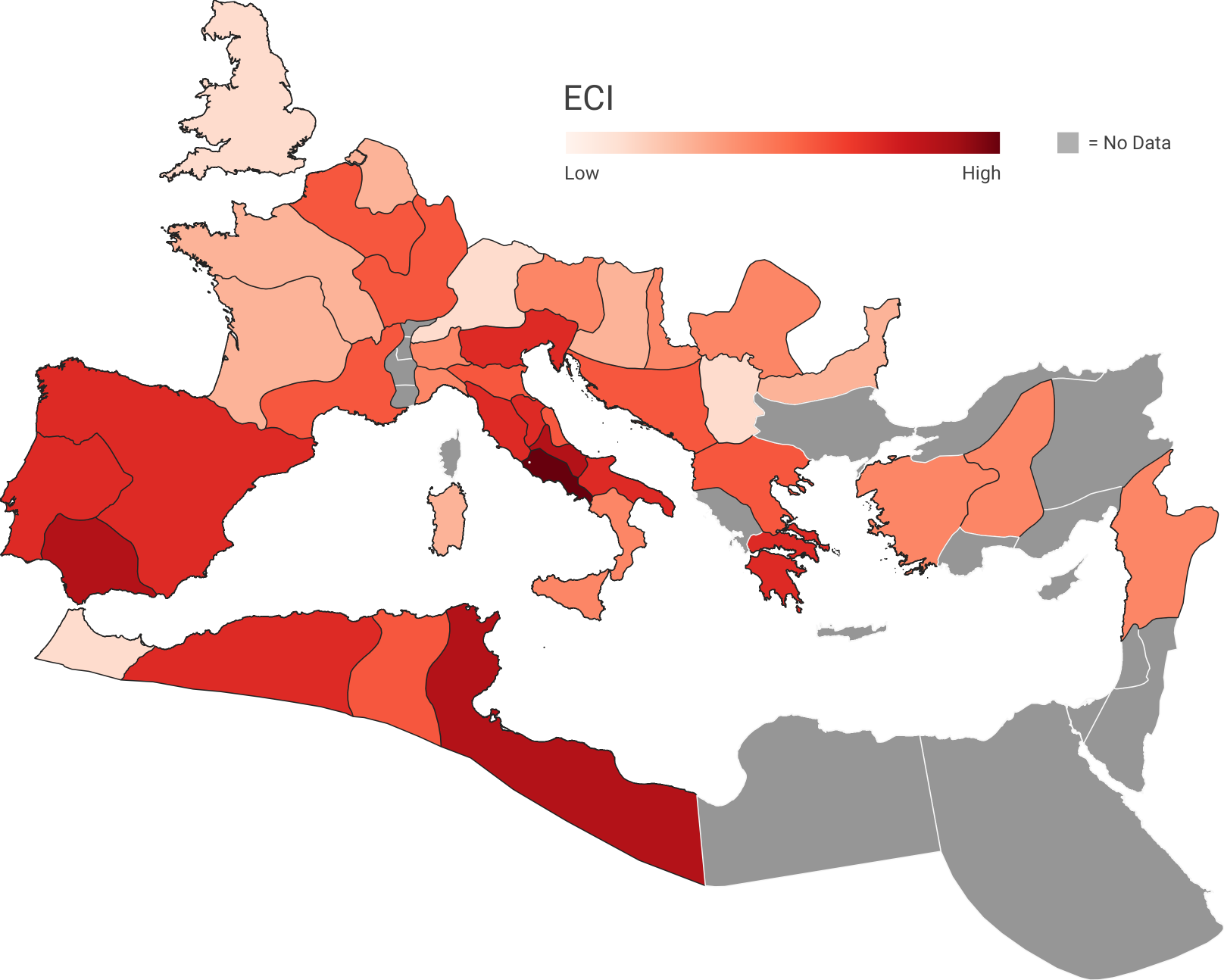}
\includegraphics[width=0.45\linewidth]{images/eci_latin.pdf}
\caption{Comparing the ECI values (color) before (left) and after (right) the bias correction.}
\label{fig:eci-bias-by-province}
\end{figure}

Figure \ref{fig:eci-bias-by-province} compares the ECI we obtain without the bias correction (left) and after the bias correction (right, which is the version we use for our main results and corresponds to Figure \ref{fig:eci-avg-by-province}). We can see there is a relative shift of complexity when we correct for biases, leading to increases in complexity in the Eastern portion of the empire and a decrease in the Western part.

\begin{figure}[t]
\centering
\includegraphics[width=\linewidth]{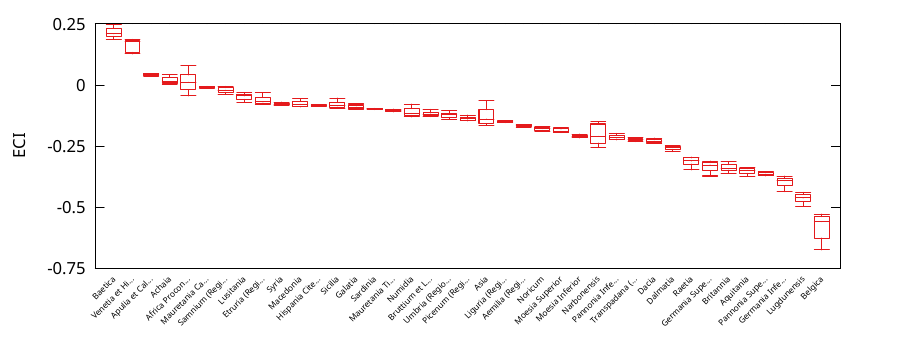}
\caption{The boxplots of ECI value distributions (y axis) per province (x axis). The boxplots show the 10th, 25th, 50th, 75th and 90th percentiles. Provinces are sorted left to right in descending order of their median ECI value. Latium is excluded.}
\label{fig:eci-boxplots}
\end{figure}

Since we draw multiple potential $M_{po}$ matrices, one question is how stable the ECI scores are across this set of possible province-occupation connections. Figure \ref{fig:eci-boxplots} shows the distributions of ECI scores for all provinces -- excluding Latium given that it is an outlier with an extremely high ECI score which does not vary much. We can see that some provinces have a higher variability than others. However, in general, the relative rankings are reasonably stable.

\subsection*{Building the Occupation Space}
The Occupation Space is a graph $G = (V, E)$. The set of nodes $V$ is the set of all $430$ occupations in our data. The set of edges $E$ should contain the most significant relationships between occupations in $V$. Here significance means that the two occupations co-appear in the same provinces more than one would expect at random.

To calculate the co-appearances, we first multiply $M_{po}$ with its own transpose: $M_{oo} = M_{po}^TM_{po}$. $M_{oo}$ gives us the raw co-appearance strength. However, in $M_{oo}$ strong connections could be due to chance: farmers and carpenters will have a strong connection only because they are very common occupations, but the significance of their relation might be lower than two rarer occupations that always co-occur together.

To fix this issue we decide to use the noise-corrected backboning technique \cite{coscia2017network}, which creates a null expectation for each pair of occupations in $M_{oo}$. We chose this technique because its null model is adequate for our data type -- being developed for discrete count edge weights which are distributed with a high skew. We then keep in $E$ only those connections that surpass the null expectation by a significant margin.

However, this creates two issues -- both of them are caused by the fact that there are several occupations that are present only in Latium. Some occupations end up not having a single connection in $E$. We fix this issue by picking their most significant connection and including it in $E$. The second issue is that the Occupation Space contains more than a single connected component. One requirement to estimate the province similarity on this network is that it must have a single connected component. Therefore, we include in $E$ also the minimum number of the most significant edges necessary to reconnect together all connected components of the network.

In both cases, we only include edges that are stronger than null expectation.

\subsection*{Economic Complexity Index}
We calculate ECI as defined in the standard approach \cite{hausmann2014atlas} of the original definition \cite{hidalgo2009building}, rather than recent evolutions \cite{hidalgo2021economic} and alternative definitions like economic fitness \cite{tacchella2012new}. Economic fitness is ill-suited for data which have strong outliers like Latium, which ends up being the only province with a non-zero complexity in that framework.

The first step is to calculate the Revealed Comparative Advantage (RCA) \cite{balassa1965trade} of each entry in $M_{po}$:

$$ M_{po} = \dfrac{M_{po} / \sum \limits_p M_{po}}{\sum \limits_o M_{po} / \sum \limits_{p,o} M_{po}}. $$

In practice, we compare the relative importance of occupation $o$ for province $p$ over the average relative importance of $o$ across all provinces. This index is equal to $1$ if $o$ is as important to $p$ as it is to the average of the empire. Values below $1$ indicate lower than average importance, and higher than $1$ a higher than average importance.

Then we binarize this transformed $M_{po}$, by equating to $1$ all $M_{po} > 1$ entries and setting zero to all other entries. In practice, we simply record the list of all $o$s for a province $p$ that are present more in $p$ than in the average Roman province.

We then produce two normalizations of this binary $M_{po}$: $P_{po} = M_{po} / \sum \limits_p M_{po}$ and $O_{po} = M_{po} / \sum \limits_o M_{po}$, which are versions of this matrix normalized by either row or column sum. The final preparation step is producing a province-province matrix:

$$ M_{pp} = O_{po}^TP_{po}.$$

$M_{pp}$ records the number of shared occupations between two provinces, normalized by two factors: how important they are for the two provinces and how common those occupations are in the empire at large.

Now the second largest eigenvector of $M_{pp}$ is the ECI measure, as it is the vector capturing the largest portion of variance in the $M_{pp}$. One way to interpret ECI is that it identifies the average province, the demarcation line between complex and simple provinces, and the vector measures how far a given province is from this border between two groups \cite{mealy2019interpreting}. Higher values mean farther on the side of complexity, while lower values indicate that the province is farther in the direction of simplicity.

The eigenvector does not have interpretable units, so it can be normalized in different ways -- as we do in the paper -- as long as the normalization is linear and does not change the ranking.

\section*{Acknowledgments}
The shapefile of Roman provinces in 200 CE (peak Roman Empire under the Severan dynasty) is based on 'roman-empire-ce-200-provinces.geojson' published by The Ancient World Mapping Centre \cite{ancient_world_mapping_centre_ancient_nodate} and corrected by Adam Pažout in 2023. 

\section*{Author Contributions}
MC and TB designed the study. PH collected the data. MM, PH, and MC wrote code and prepared the figures. All authors wrote and approved the manuscript.

\section*{Competing interests}
The authors declare no competing interests.

\section*{Data availability}
All data is available. The inscription data is available on Zenodo at DOIs 10.5281/zenodo.10473706 and 10.5281/zenodo.10139110. The trade data is available in the Harvard Dataverse at DOIs 10.7910/DVN/H8SFD2\&version=10.0 and 10.7910/DVN/3BAL1O\&version=6.0. Code and additional data available on our Github: https://github.com/past-networks/roman\_ECI/.

\section*{Funding statement}
This research was conducted within the framework of the Past Social Networks Project (2023-2026), funded by The Carlsberg Foundation’s Young Researcher Fellowship (CF21-0382). This work was supported by a research grant (VIL57402) from VILLUM FONDEN.

\bibliography{ref}

\end{document}